\documentclass[12pt,titlepage,letterpaper]{article}
\usepackage{amsmath}
\usepackage{amssymb}
\usepackage{amsfonts}
\usepackage{charter}
\usepackage{eucal}
\usepackage[onehalfspacing]{setspace}
\usepackage{graphicx}
\usepackage[top=1.5in,bottom=1in]{geometry}

\newtheorem{theorem}{Theorem}

\input{tcilatex}

\begin{document}

\title{The functional singular value decomposition for bivariate stochastic
processes}
\author{Daniel Gervini\thanks{%
Department of Mathematical Sciences, University of Wisconsin--Milwaukee,
P.O. Box 413, Milwaukee, WI 53201 (email: gervini@uwm.edu).} \\
%EndAName
University of Wisconsin--Milwaukee}
\maketitle

\begin{abstract}
In this article we present some statistical applications of the functional
singular value decomposition (FSVD). This tool allows us to decompose the
sample mean of a bivariate stochastic process into components that are
functions of separate variables. These components are sometimes
interpretable functions that summarize salient features of the data. The
FSVD can be used to visually detect outliers, to estimate the mean of a
stochastic process or to obtain individual smoothers of the sample surfaces.
As estimators of the mean, we show by simulation that FSVD estimators are
competitive with tensor-product splines in some situations.

\end{abstract}

\section{Introduction}

The analysis of samples of curves has become more common in statistical
applications in recent years. In many applications, the data consists of
discrete realizations of a univariate process, say $X(t)$, where $t$ can be
time (e.g.~growth curves in Gasser et al., 2004), distance (e.g.~biomarker
expression curves in Morris and Carroll, 2006) or age (e.g.~income
distribution densities in Kneip and Utikal, 2001), among other
possibilities. More examples and statistical methodology can be found in
Ramsay and Silverman (2002, 2005) or Ferraty and Vieu (2006).

Multivariate stochastic processes, on the other hand, have received less
attention. By multivariate process we mean a real-valued process $X(\mathbf{s%
})$ that is a function of a multidimensional variable $\mathbf{s}$. They are
also known as random fields (Adler and Taylor, 2007). Although they are less
common in statistics than univariate processes, they play an important role
in fMRI studies and spatial statistics (Taylor and Worseley, 2007; Nychka,
2000). In these applications $\mathbf{s}$ is a point in $\mathbb{R}^{2}$ or $%
\mathbb{R}^{3}$. However, in other situations the variables do not belong to
a single natural space. For example, $X(s,t)$ may be the mortality rate for
individuals of age $s$ during year $t$ in a given country, or the outcome of
a multichannel electroencephalography study where $t$ is time and $s$ is the
location of the electrode on the scalp. It is clear that the variables $s$
and $t$ belong to different spaces; although the product space could be
regarded as a single space, this would be more a mathematical formalization
than a natural structure implied by the data.

\FRAME{ftbpFU}{6.0277in}{4.4884in}{0pt}{\Qcb{Human Mortality Data. Mean of
log-mortality rates for ten European countries.}}{\Qlb{fig:Raw_Mean}}{%
rawmean.pdf}{\special{language "Scientific Word";type
"GRAPHIC";maintain-aspect-ratio TRUE;display "ICON";valid_file "F";width
6.0277in;height 4.4884in;depth 0pt;original-width 10.8102in;original-height
8.0376in;cropleft "0";croptop "1";cropright "1";cropbottom "0";filename
'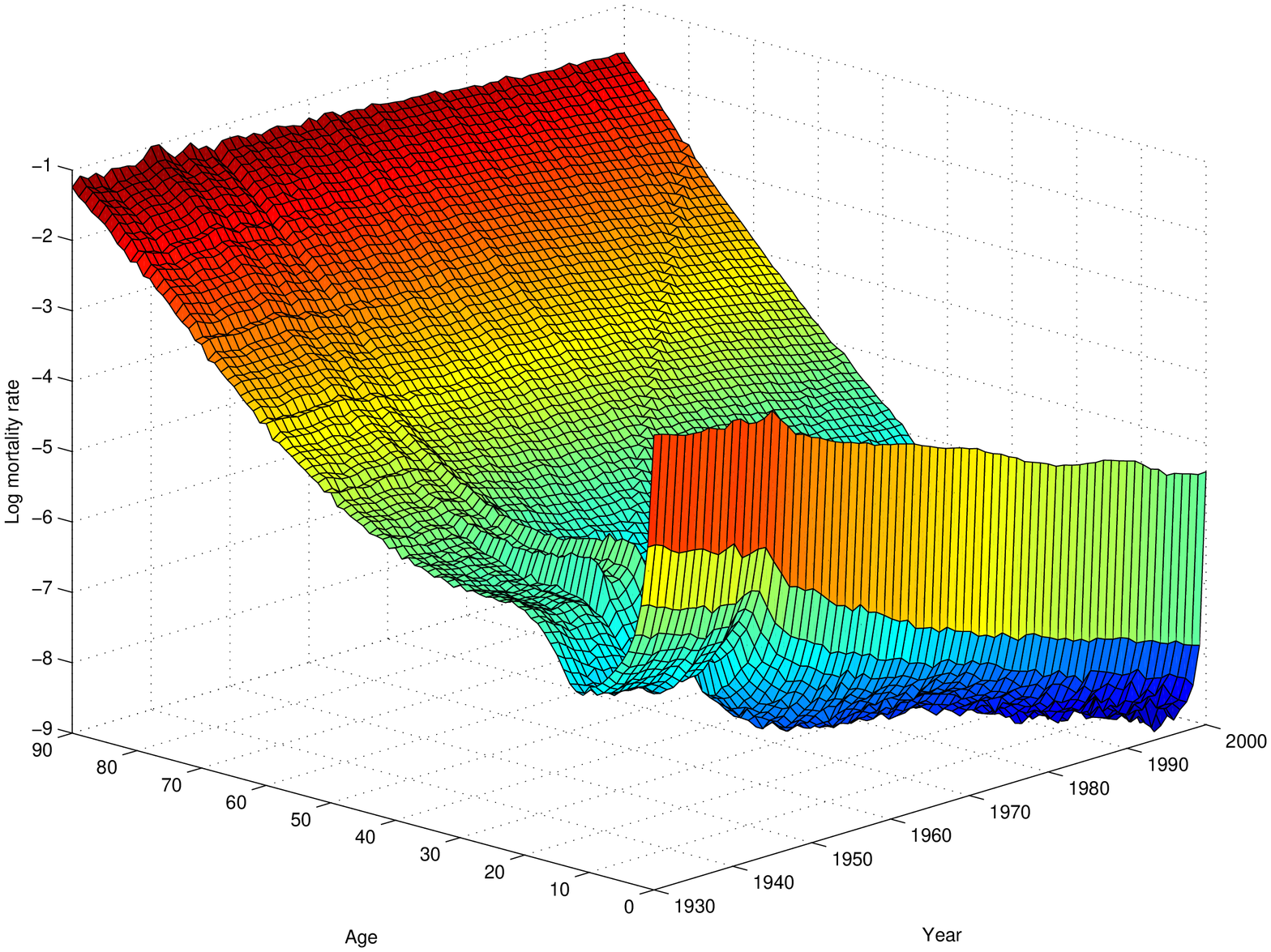';file-properties "XNPEU";}}

To understand more clearly the problems involved, in Fig.~\ref{fig:Raw_Mean}
we have plotted the sample mean of log-mortality rates for ten European
countries. The raw mean shows some irregularities due to random noise. To
regularize a bivariate estimator like this, one would normally employ a
smoothing method based on splines (Gu, 2000) or kernels (H\"{a}rdle and M%
\"{u}ller, 2000). However, those global smoothers will most likely level off
important features of the data, like the increased mortality rates during
the Second World War, which are sharp but localized features.

In this paper we present a different approach, based on a generalization of
the singular value decomposition. The basic idea is to approximate a
bivariate function $\mu (s,t)$ with a sum of functions of separate
variables, $\mu ^{(p)}(s,t)=\sum_{k=1}^{p}\lambda _{k}^{1/2}\phi _{k}(s)\psi
_{k}(t)$, where $\phi _{k}$ and $\psi _{k}$ are univariate functional
principal components (Silverman, 1996; Yao and Lee, 2006; Gervini, 2006).
The components are sometimes interpretable functions that summarize
important features of the data, and can be used, for example, to detect
atypical observations. Individual smoothers of the sample surfaces can also
be obtained as by-products. The bivariate singular value decomposition has
been used in image analysis and physics (Dente et al., 1996; Aubry et al.,
1991), under the name of \textquotedblleft biorthogonal
decomposition\textquotedblright .\ However, these articles disregard
smoothing issues, using raw principal components for estimation. In most
statistical applications, that would lead to extremely noisy and
uninformative estimates. In contrast, the method we present here produces
smooth and regular estimators.

This article is organized as follows. The functional singular value
decomposition (FSVD) is presented in Section \ref{sec:FSVD}, and smooth
estimators of the components are introduced in Section \ref{sec:Estimation}.
An application to a real dataset in Section\ \ref{sec:Example} illustrates
the potential of the FSVD as a graphical tool. In Section\ \ref%
{sec:Simulations} we compare by simulation the behavior of the FSVD with
tensor-product splines as estimators of the mean. Abbreviated proofs of the
theorems are given in the Appendix; more detailed proofs and additional
material is available on a Technical Report that will be posted on the
author's website.

\section{The functional singular value decomposition\label{sec:FSVD}}

Let $X(s,t)$ be a real-valued stochastic process in $L^{2}(\mathcal{S}\times 
\mathcal{T})$ with finite expectation $\mu (s,t)$ and finite covariance
function $\rho \{(s_{1},t_{1}),(s_{2},t_{2})\}$. We assume that $\mathcal{S}$
and $\mathcal{T}$ are closed intervals in $\mathbb{R}$. Let us define the
kernel functions 
\begin{equation*}
k_{1}(s_{1},s_{2})=\int_{\mathcal{T}}\mu (s_{1},t)\mu (s_{2},t)\ dt
\end{equation*}%
and 
\begin{equation*}
k_{2}(t_{1},t_{2})=\int_{\mathcal{S}}\mu (s,t_{1})\mu (s,t_{2})\ ds.
\end{equation*}%
We say that $\phi \in L^{2}(\mathcal{S})$ is an eigenfunction of $k_{1}$
with eigenvalue $\lambda $ if $\int_{\mathcal{S}}k_{1}(s,u)\phi
(u)du=\lambda \phi (s)$ for almost every $s\in \mathcal{S}$. The
eigenfunctions of $k_{2}$ are defined in a similar way, only that they
belong to $L^{2}(\mathcal{T})$. The next theorem establishes the existence
of a decomposition of $k_{1}$, $k_{2}$ and $\mu $ in terms of these
eigenfunctions.

\begin{theorem}
\label{thm:FSVD}There exist a non-increasing sequence of positive
eigenvalues $\{\lambda _{k}\}$ of $k_{1}$ and $k_{2}$, an orthonormal
sequence $\{\phi _{k}\}$ of eigenfunctions of $k_{1}$ and an orthonormal
sequence $\{\psi _{k}\}$ of eigenfunctions of $k_{2}$ such that 
\begin{equation}
k_{1}(s_{1},s_{2})=\sum_{k\geq 1}\lambda _{k}\phi _{k}(s_{1})\phi
_{k}(s_{2}),  \label{eq:k1_expansion}
\end{equation}%
\begin{equation}
k_{2}(t_{1},t_{2})=\sum_{k\geq 1}\lambda _{k}\psi _{k}(t_{1})\psi _{k}(t_{2})
\label{eq:k2_expansion}
\end{equation}%
and 
\begin{equation}
\mu (s,t)=\sum_{k\geq 1}\lambda _{k}^{1/2}\phi _{k}(s)\psi _{k}(t).
\label{eq:mu_expansion}
\end{equation}%
The series (\ref{eq:k1_expansion}), (\ref{eq:k2_expansion}) and (\ref%
{eq:mu_expansion}) converge in the sense of the $L^{2}$ norm. If in addition 
$\mu (s,t)$ is continuous, then $\{\phi _{k}\}$ and $\{\psi _{k}\}$ are
continuous functions and the convergence of (\ref{eq:k1_expansion}) and (\ref%
{eq:k2_expansion}) is absolute and uniform in both variables, with the
identities holding for each $(s_{1},s_{2})$ and each $(t_{1},t_{2})$. If the
right-hand side of (\ref{eq:mu_expansion}) converges uniformly and
absolutely, then the identity also holds for every $(s,t)$.
\end{theorem}

Theorem \ref{thm:FSVD} implies that the truncated series 
\begin{equation}
\mu ^{(p)}(s,t)=\sum_{k=1}^{p}\lambda _{k}^{1/2}\phi _{k}(s)\psi _{k}(t)
\label{eq:mu_p}
\end{equation}%
converges to $\mu (s,t)$ in the sense of $L^{2}(\mathcal{S}\times \mathcal{T}%
)$ as $p$ increases, and that the convergence is pointwise for every $(s,t)$
if the right-hand side of (\ref{eq:mu_expansion}) converges uniformly and
absolutely. The latter occurs if, for instance, the $\phi _{k}$s and the $%
\psi _{k}$s are uniformly bounded and $\sum_{k\geq 1}\lambda _{k}^{1/2}$ is
finite.

In analogy with the multivariate singular value decomposition, the truncated
series $\mu ^{(p)}$ given by (\ref{eq:mu_p}) provides the best possible
approximation of $\mu $ among linear combinations of functions of separate
variables, in the sense of the $L^{2}(\mathcal{S}\times \mathcal{T})$ norm.

\begin{theorem}
\label{thm:best_approx}Let $\mathcal{H}_{p}$ be the class of functions $%
h(s,t)=\sum_{k=1}^{p}a_{k}f_{k}(t)g_{k}(s)$ with $\{f_{k}\}$ and $\{g_{k}\}$
orthonormal in $L^{2}(\mathcal{T})$ and $L^{2}(\mathcal{S})$, respectively.
Then 
\begin{equation*}
\min_{h\in \mathcal{H}_{p}}\left\Vert \mu -h\right\Vert ^{2}=\Vert \mu -\mu
^{(p)}\Vert ^{2},
\end{equation*}%
with $\mu ^{(p)}$ as in (\ref{eq:mu_p}).
\end{theorem}

The function $\mu ^{(p)}(s,t)$ is the sum of $p$ functions of separate
variables, $d_{k}(s,t)=\lambda _{k}^{1/2}\phi _{k}(s)\psi _{k}(t)$, that we
will call \textquotedblleft detail functions\textquotedblright .\ The detail
functions are orthogonal in both variables, and $\left\Vert d_{k}\right\Vert
=\lambda _{k}^{1/2}$, so they provide finer levels of detail as $k$
increases. An appealing feature of the detail functions is that they are
often interpretable functions, giving us information about the most relevant
characteristics of the process under investigation.

Of course, all this would be of little practical use if the computation of
the $\phi _{k}$s and $\psi _{k}$s required a good preliminary estimator of $%
\mu $. But we show below that good estimators of the eigenfunctions can be
obtained from the raw data, and these estimators are then used to construct
a smooth estimator of $\mu $.

\section{Smooth estimation of the eigenfunctions\label{sec:Estimation}}

Let $X_{1},\ldots ,X_{n}$ be an i.i.d.~sample of the process $X$. In most
cases, the $X_{i}$s are observed on a discrete grid $\{s_{j}\}\times
\{t_{k}\}\subset \mathcal{S}\times \mathcal{T}$ with random error, so the
data follows the model 
\begin{equation}
x_{ijk}=X_{i}(s_{j},t_{k})+\varepsilon _{ijk},\ i=1,\ldots ,n,~j=1,\ldots
,m,~k=1,\ldots ,r.  \label{eq:raw_model}
\end{equation}%
We will assume that $\mathrm{E}(\varepsilon _{ijk})=0$, $\varepsilon _{ijk}$
is independent of $X_{i}$, $\varepsilon _{ijk}$ and $\varepsilon _{i^{\prime
}j^{\prime }k^{\prime }}$ are independent if $i\neq i^{\prime }$, and $%
\mathrm{E}(\varepsilon _{ijk}\varepsilon _{ij^{\prime }k^{\prime }})=\sigma
^{2}\delta _{jj^{\prime }}\delta _{kk^{\prime }}$ (where $\delta $ is
Kronecker's delta).

The simplest estimator of $\mu $ at the grid points is the cross sectional
mean, $\hat{\mu}(s_{j},t_{k})=\sum_{i=1}^{n}x_{ijk}/n$. The corresponding
estimators of the kernel functions $k_{1}$ and $k_{2}$, using the trapezoid
rule for numerical integration, are 
\begin{equation*}
\hat{k}_{1}(s_{j},s_{j^{\prime }})=\sum_{k=1}^{r}u_{k}\hat{\mu}(s_{j},t_{k})%
\hat{\mu}(s_{j^{\prime }},t_{k})
\end{equation*}%
and 
\begin{equation*}
\hat{k}_{2}(t_{k},t_{k^{\prime }})=\sum_{j=1}^{m}v_{j}\hat{\mu}(s_{j},t_{k})%
\hat{\mu}(s_{j},t_{k^{\prime }}),
\end{equation*}%
where $u_{1}=(t_{2}-t_{1})/2$, $u_{k}=(t_{k+1}-t_{k-1})/2$, $k=2,\ldots ,r-1$%
, $u_{r}=(t_{r}-t_{r-1})/2$, and $v_{1}=(s_{2}-s_{1})/2$, $%
v_{j}=(s_{j+1}-s_{j-1})/2$, $j=2,\ldots ,m-1$, $v_{m}=(s_{m}-s_{m-1})/2$.

From $\hat{k}_{1}$ and $\hat{k}_{2}$ we can compute smooth estimators of the
eigenfunctions $\{\phi _{k}\}$ and $\{\psi _{k}\}$ using spline models (such
as B-splines; de Boor, 2001) as follows. We know that 
\begin{equation*}
\phi _{1}=\func{argmax}_{\left\Vert g\right\Vert =1}\tiint
k_{1}(s_{1},s_{2})g(s_{1})g(s_{2})\mathrm{d}s_{1}\mathrm{d}s_{2}.
\end{equation*}%
Then, given a spline basis $\{\beta _{1},\ldots ,\beta _{q}\}$ in $L^{2}(%
\mathcal{S})$, we write $g(s)=\sum_{j=1}^{q}b_{j}\beta _{j}(s)$ and define 
\begin{equation*}
\mathbf{\hat{b}}_{1}=\func{argmax}\{\mathbf{b}^{T}\mathbf{\hat{\Omega}b}:%
\mathbf{b}^{T}\mathbf{\Gamma b}=1\},
\end{equation*}%
where $\hat{\Omega}_{ij}=\iint \hat{k}_{1}(s_{1},s_{2})\beta
_{i}(s_{1})\beta _{j}(s_{2})\mathrm{d}s_{1}\mathrm{d}s_{2}$ and $\Gamma
_{ij}=\int \beta _{i}(s)\beta _{j}(s)\mathrm{d}s$. Then $\hat{\phi}%
_{1}(s)=\sum_{j=1}^{q}\hat{b}_{1j}\beta _{j}(s)$ is a spline estimator of
the first eigenfunction of $k_{1}$.

For the rest of the eigenfunctions we proceed sequentially: since 
\begin{equation*}
\phi _{k}=\func{argmax}\left\{ \tiint k_{1}(s_{1},s_{2})g(s_{1})g(s_{2})%
\mathrm{d}s_{1}\mathrm{d}s_{2}:\left\Vert g\right\Vert =1\text{ and }\langle
g,\phi _{j}\rangle =0\text{ for }j<k\right\} ,
\end{equation*}%
we define 
\begin{equation}
\mathbf{\hat{b}}_{k}=\func{argmax}\{\mathbf{b}^{T}\mathbf{\hat{\Omega}b}:%
\mathbf{b}^{T}\mathbf{\Gamma b}=1,\mathbf{b}^{T}\mathbf{\Gamma \hat{b}}%
_{j}=0,j<k\}  \label{eq:b_k_hat}
\end{equation}%
and set $\hat{\phi}_{k}(s)=\sum_{j=1}^{q}\hat{b}_{kj}\beta _{j}(s)$. The
corresponding eigenvalues can be estimated by $\hat{\lambda}_{k}=\mathbf{%
\hat{b}}_{k}^{T}\mathbf{\hat{\Omega}\hat{b}}_{k}$.

Computationally, (\ref{eq:b_k_hat}) is a very simple problem. Let $\mathbf{V}%
=\mathrm{diag}(v_{1},\ldots ,v_{m})$, $\mathbf{B}\in \mathbb{R}^{q\times m}$
with $B_{ij}=\beta _{i}(s_{j})$, and $\mathbf{K}_{1}\in \mathbb{R}^{m\times
m}$ with $K_{1ij}=\hat{k}_{1}(s_{i},s_{j})$. Then, using the trapezoid rule
for numerical integration, $\mathbf{\hat{\Omega}=B}^{T}\mathbf{VK}_{1}%
\mathbf{VB}$ and $\mathbf{\Gamma =B}^{T}\mathbf{VB}$. If $\mathbf{\Gamma }%
^{1/2}$ denotes the symmetric square root of $\mathbf{\Gamma }$ and $\mathbf{%
\hat{c}}_{k}$ the $k$th unit-norm eigenvector of $\mathbf{\Gamma }^{-1/2}%
\mathbf{\hat{\Omega}\Gamma }^{-1/2}$, then $\mathbf{\hat{b}}_{k}\mathbf{%
=\Gamma }^{-1/2}\mathbf{\hat{c}}_{k}$.

If the true eigenfunctions belong to the space generated by the specified
spline basis, and the eigenvalues of $\mathbf{\Gamma }^{-1/2}\mathbf{\Omega
\Gamma }^{-1/2}$ (with $\mathbf{\Omega }$ given below) have multiplicity
one, then the above estimators are consistent. This is a consequence of the
next theorem together with the results of Tyler (1981).

\begin{theorem}
\label{thm:Consistency}Let $\mathbf{\Omega }\in \mathbb{R}^{q\times q}$ be
given by $\Omega _{ij}=\iint k_{1}(s_{1},s_{2})\beta _{i}(s_{1})\beta
_{j}(s_{2})\mathrm{d}s_{1}\mathrm{d}s_{2}$. If $\max v_{j}\rightarrow 0$ as $%
m\rightarrow \infty $ and $\max u_{k}\rightarrow 0$ as $r\rightarrow \infty $%
, then $\mathbf{\hat{\Omega}}\rightarrow \mathbf{\Omega }$ in probability as 
$n$, $m$ and $r$ go to infinity.
\end{theorem}

In practice, though, the eigenfunctions may not belong to a spline space.
But the asymptotic bias will be negligible if the spline basis is
appropriately chosen. For that reason, in this paper we use adaptive
free-knot splines as in Gervini (2006). Another possibility is to use a
large number of basis functions with global regularization, as in Silverman
(1996), but we prefer the free-knot approach because it provides better fits
for the local features of the eigenfunctions.

Concretely, the algorithm we implemented aggregates knots by maximizing (\ref%
{eq:b_k_hat}) over a grid of candidates (usually the grid $\{s_{j}\}$
itself) until there is no significant improvement on the objective function (%
\ref{eq:b_k_hat}). Repeated knots are allowed, since they provide better
resolution of the local features of the components (at the expense of fewer
degrees of differentiability). The optimal number of knots can be chosen
either subjectively or by cross-validation. This procedure must be repeated
for each component because the optimal placement and number of knots changes
with each component.

The eigenfunctions $\{\psi _{k}\}$ of $k_{2}$ are estimated in a similar
way, using a spline basis in $L^{2}(\mathcal{T})$. Since the choice of sign
of the eigenfunctions is always arbitrary, care must be taken so that $\hat{%
\lambda}_{k}^{1/2}=\iint \hat{\mu}(s,t)\hat{\phi}_{k}(s)\hat{\psi}_{k}(t)%
\mathrm{d}s\mathrm{d}t$ is positive. As before, we use the trapezoid rule
for numerical integration, so $\hat{\lambda}_{k}^{1/2}=\hat{\phi}_{k}(%
\mathbf{s})^{T}\mathbf{V\bar{X}U}\hat{\psi}_{k}(\mathbf{t})$, where $\hat{%
\phi}_{k}(\mathbf{s})$ is the vector with elements $\hat{\phi}_{k}(s_{j})$
and $\hat{\psi}_{k}(\mathbf{t})$ is the vector with elements $\hat{\psi}%
_{k}(t_{j})$; $\mathbf{\bar{X}}$ is the average of the matrices $\mathbf{X}%
_{i}$ with elements $(X_{i})_{jk}=x_{ijk}$ and $\mathbf{U}=\mathrm{diag}%
(u_{1},\ldots ,u_{r})$.

The eigenfunctions are estimated sequentially until a given order $p$, and
then we define 
\begin{equation*}
\hat{\mu}^{(p)}(s,t)=\sum_{k=1}^{p}\hat{\lambda}_{k}^{1/2}\hat{\phi}_{k}(s)%
\hat{\psi}_{k}(t).
\end{equation*}%
The order $p$ must be chosen with care, to reduce bias as much as possible.
For reasons that will become clearer in Sections\ \ref{sec:Example} and \ref%
{sec:Simulations}, we recommend to use a large $p$ as long as the estimators
of the eigenfunctions are not overwhelmed by noise, even if the
corresponding $\hat{\lambda}_{k}$s seem to be negligibly small.

Interestingly, $\hat{\mu}^{(p)}$ can be further decomposed into terms that
represent the individual contributions of the $X_{i}$s, since $\hat{\lambda}%
_{k}^{1/2}=\sum_{i=1}^{n}\hat{w}_{ik}/n$ with $\hat{w}_{ik}=\hat{\phi}_{k}(%
\mathbf{s})^{T}\mathbf{VX}_{i}\mathbf{U}\hat{\psi}_{k}(\mathbf{t})$. Note
that $\hat{w}_{ik}$ is an estimator of $w_{ik}=\iint X_{i}(s,t)\phi
_{k}(s)\psi _{k}(t)\mathrm{d}s\mathrm{d}t$. Then we can define individual
predictors of the unobserved sample paths $X_{i}(s,t)$, 
\begin{equation*}
\hat{X}_{i}^{(p)}(s,t)=\sum_{k=1}^{p}\hat{w}_{ik}\hat{\phi}_{k}(s)\hat{\psi}%
_{k}(t).
\end{equation*}%
The score vectors $\mathbf{\hat{w}}_{i}$ are useful for exploratory data
analysis; for example, they may reveal outliers or unusual groupings in the
data, as we show by example in Section\ \ref{sec:Example}. The predictors $%
\hat{X}_{i}^{(p)}$ can also be used to select the best order $p$ by
cross-validation.

\section{Example: evolution of human mortality in the 20th century\label%
{sec:Example}}

The socioeconomic progress experienced by western European countries after
the Second World War is very graphically exemplified by the evolution of
human mortality curves. Mortality rates, which are the percentages of people
of certain age who die in a given year, can be seen as longitudinal of
functional data in two senses: for a given year, mortality rates are a
function of age; and for each age, the evolution of mortality rates over the
years are a time series. But a thorough statistical analysis must take into
account the interplay between these two variables; that is, the data must be
seen as realizations of a bivariate stochastic process.

In this section we analyze mortality rates between the years of 1930 and
2000, for people ranging from 0 to 90 years of age. The data was downloaded
from the Human Mortality Database website, www.mortality.org. We only
included countries of western Europe for which complete data was available:
Belgium, Denmark, England, Finland, France, Italy, the Netherlands, Norway,
Spain and Sweden. For country $i$ we defined $X_{i}(s,t)$ as the logarithm
of the mortality rate for age $s$ at year $t$; the data was observed on the
grid $\{0,1,\ldots ,90\}\times \{1930,1931,\ldots ,2000\}$.

\FRAME{ftbpFU}{4.9372in}{4.4901in}{0pt}{\Qcb{Human Mortality Data. Free-knot
spline estimators of the eigenfunctions: (a) $\hat{\protect\phi}_{1}(s)$,
(b) $\hat{\protect\psi}_{1}(t)$, (c) $\hat{\protect\phi}_{2}(s)$, (d) $\hat{%
\protect\psi}_{2}(t)$, (e) $\hat{\protect\phi}_{3}(s)$ and (f) $\hat{\protect%
\psi}_{3}(t)$.}}{\Qlb{fig:eigenfunctions}}{eigenfunctions.eps}{\special%
{language "Scientific Word";type "GRAPHIC";maintain-aspect-ratio
TRUE;display "ICON";valid_file "F";width 4.9372in;height 4.4901in;depth
0pt;original-width 9.7914in;original-height 8.8998in;cropleft "0";croptop
"1";cropright "1";cropbottom "0";filename
'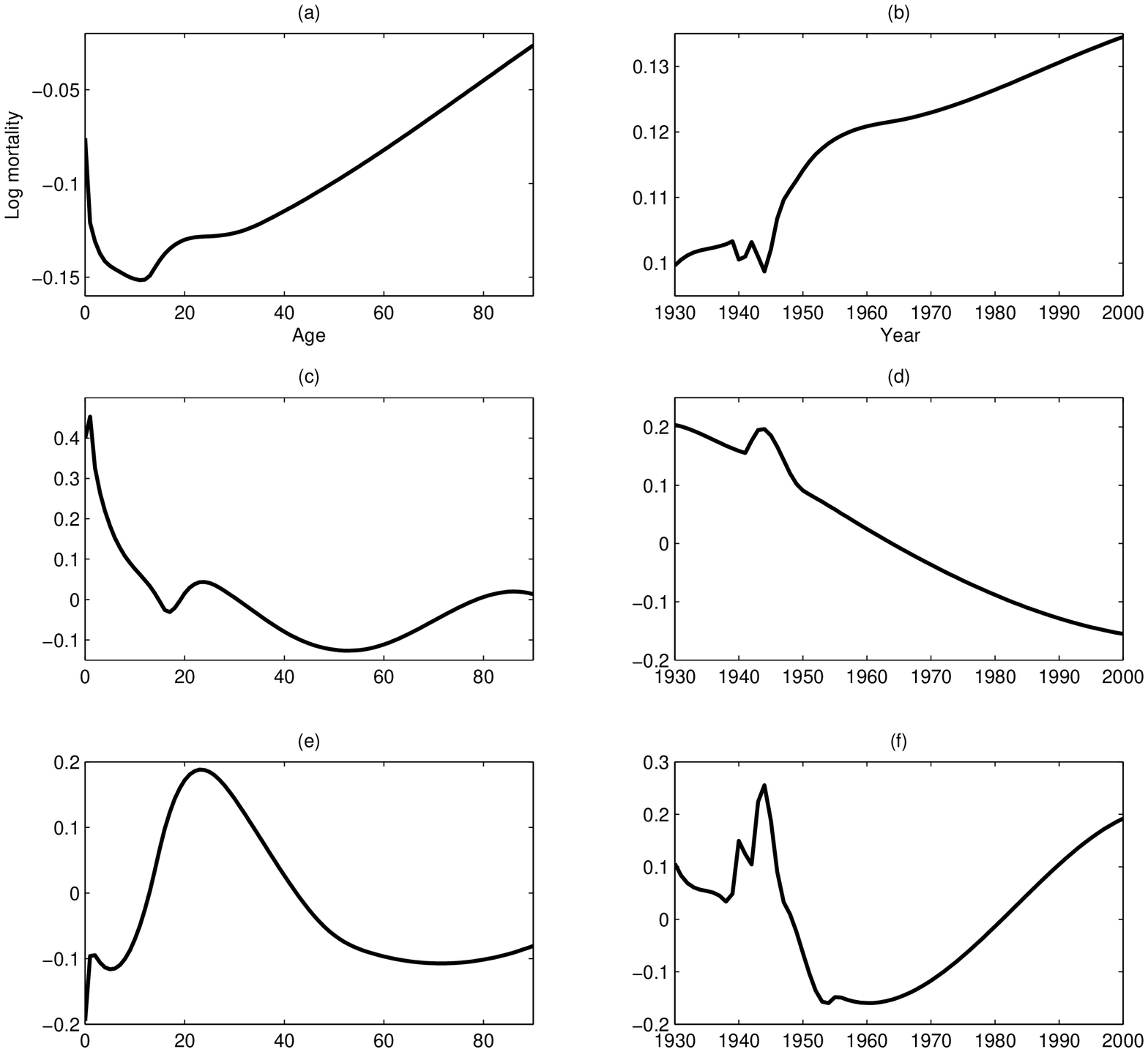';file-properties "XNPEU";}}

We computed three pairs of eigenfunctions, which are shown in Fig.~\ref%
{fig:eigenfunctions}. The corresponding root-eigenvalues were $\hat{\lambda}%
_{1}^{1/2}=435.85$, $\hat{\lambda}_{2}^{1/2}=11.09$ and $\hat{\lambda}%
_{3}^{1/2}=6.71$. Clearly, the first eigenvalue is dominant. However, the
second and third detail functions do improve the fit in ways that are
visually noticeable (the fact that obvious visual improvements may be
associated with very small eigenvalues was observed by Dente et al., 1996).

We see that $\hat{\phi}_{1}(s)$ (Fig.~\ref{fig:eigenfunctions}(a)) can be
interpreted as the basic shape of a human mortality curve: high infant
mortality is followed by a sharp decrease until adolescence, then a sharp
increase occurs that levels off at ages 20 to 30, followed by a steady
increase from then on. The companion eigenfunction $\hat{\psi}_{1}(t)$ (Fig.~%
\ref{fig:eigenfunctions}(b)) is the overall mortality trend over this
71-year period: a modest decrease in the early 30's was punctuated by the
Second World War, followed by a remarkably fast decrease in mortality that
has continued until these days. The first-order approximation $\hat{\mu}%
^{(1)}$ is depicted in Fig.~\ref{fig:raw_mu1}, together with the raw mean.
We see that the approximation is very good, but some flaws are obvious. For
example, newborn mortality ($s=0$) remains constant over the years in Fig.~%
\ref{fig:raw_mu1}(b) while it is obviously decreasing in Fig.~\ref%
{fig:raw_mu1}(a).

The second component $\hat{\phi}_{2}(s)$ (Fig.~\ref{fig:eigenfunctions}(c))
is mostly related to infant mortality, with $\hat{\psi}_{2}(t)$ (Fig.~\ref%
{fig:eigenfunctions}(d)) showing a steady decrease over the years except for
the war period. Clearly, $\hat{\mu}^{(2)}$ (Fig. \ref{fig:d2_mu2}(b))
provides a better fit for infant mortality than $\hat{\mu}^{(1)}$. The
third-order approximation $\hat{\mu}^{(3)}$ (Fig.~\ref{fig:d3_mu3}(b))
improves the fit for the war years. Note that for this period, $\hat{\mu}%
^{(2)}$ underestimates mortality for ages 20 to 30 and overestimates it for
ages 60 and over. Higher levels of detail could be added, but it is hard to
see any features of the raw mean that have not been accounted for by $\hat{%
\mu}^{(3)}$.

\FRAME{ftbpFU}{7.184in}{3.1073in}{0pt}{\Qcb{Human Mortality Data. (a) Raw
mean and (b) first-order singular value approximation.}}{\Qlb{fig:raw_mu1}}{%
raw_mu1.pdf}{\special{language "Scientific Word";type
"GRAPHIC";maintain-aspect-ratio TRUE;display "ICON";valid_file "F";width
7.184in;height 3.1073in;depth 0pt;original-width 13.6139in;original-height
5.8591in;cropleft "0";croptop "1";cropright "1";cropbottom "0";filename
'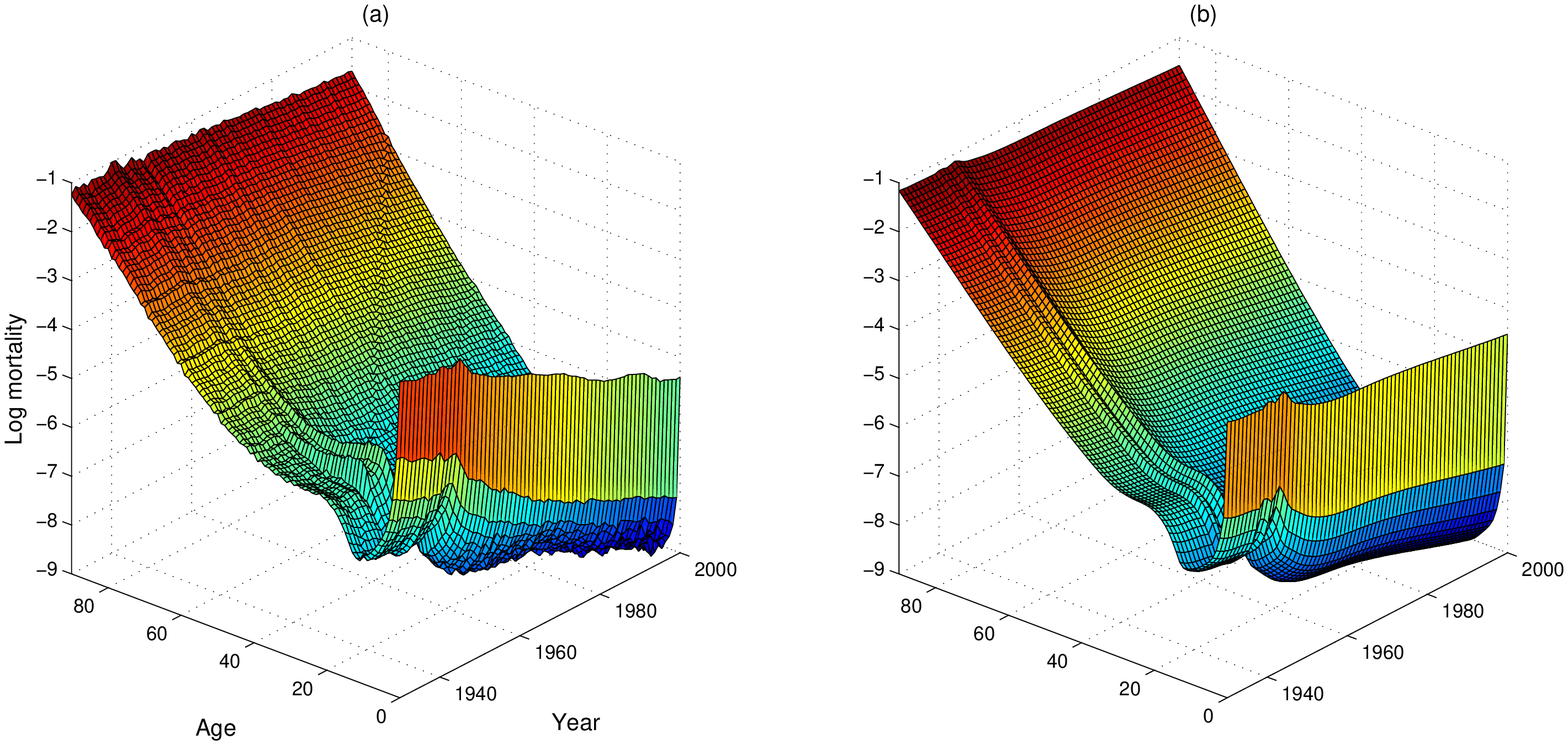';file-properties "XNPEU";}}

\FRAME{ftbpFU}{7.184in}{3.1073in}{0pt}{\Qcb{Human Mortality Data. (a)
Second-order detail function and (b) second-order singular value
approximation of the mean.}}{\Qlb{fig:d2_mu2}}{d2_mu2.pdf}{\special{language
"Scientific Word";type "GRAPHIC";maintain-aspect-ratio TRUE;display
"ICON";valid_file "F";width 7.184in;height 3.1073in;depth 0pt;original-width
13.6139in;original-height 5.8591in;cropleft "0";croptop "1";cropright
"1";cropbottom "0";filename '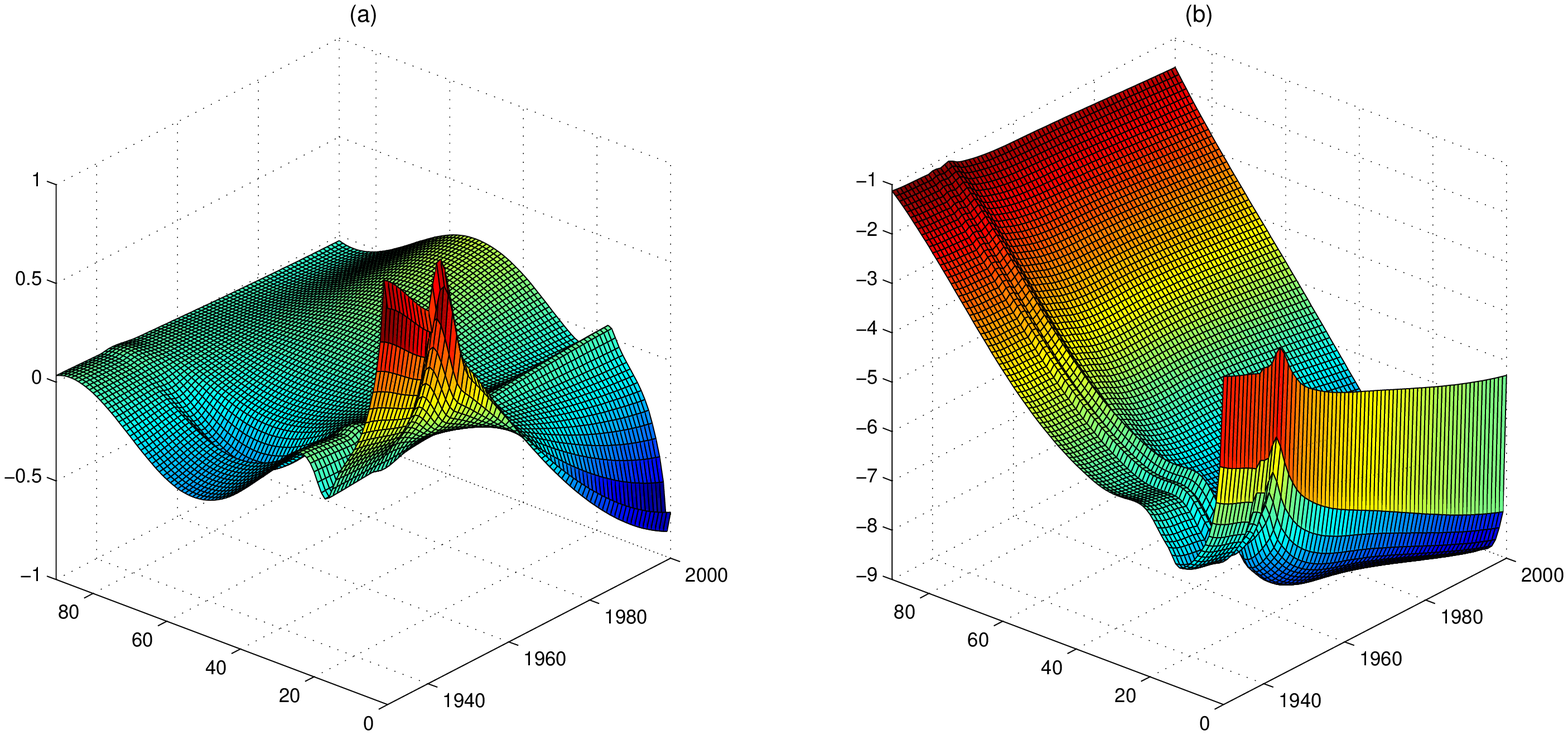';file-properties "XNPEU";}}

\FRAME{ftbpFU}{7.184in}{3.1073in}{0pt}{\Qcb{Human Mortality Data. (a)
Third-order detail function and (b) third-order singular value approximation
of the mean.}}{\Qlb{fig:d3_mu3}}{d3_mu3.pdf}{\special{language "Scientific
Word";type "GRAPHIC";maintain-aspect-ratio TRUE;display "ICON";valid_file
"F";width 7.184in;height 3.1073in;depth 0pt;original-width
13.6139in;original-height 5.8591in;cropleft "0";croptop "1";cropright
"1";cropbottom "0";filename '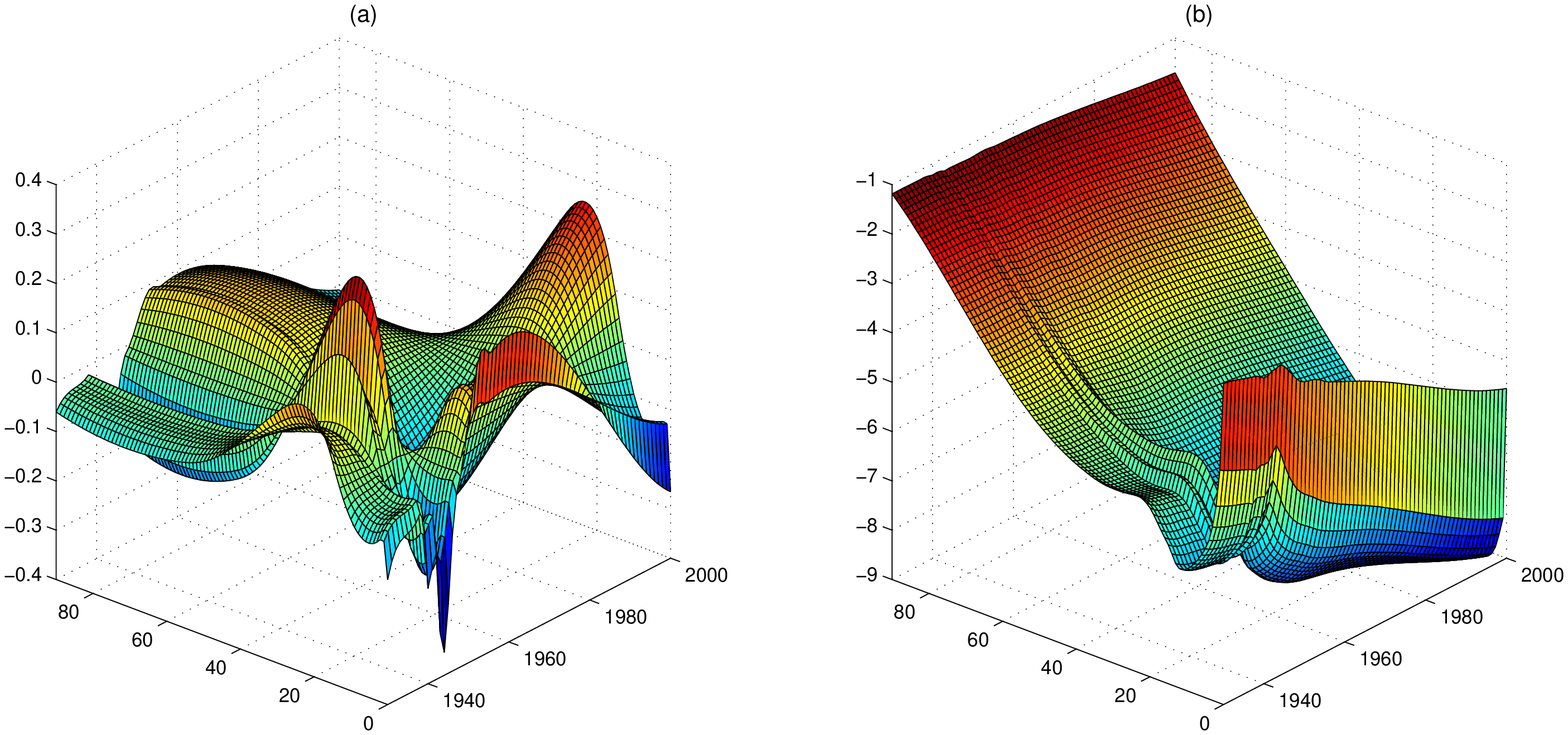';file-properties "XNPEU";}}

\FRAME{ftbpFU}{3.5423in}{2.6403in}{0pt}{\Qcb{Human Mortality Data.
Individual component scores of the ten countries.}}{\Qlb{fig:scores}}{%
scores.eps}{\special{language "Scientific Word";type
"GRAPHIC";maintain-aspect-ratio TRUE;display "ICON";valid_file "F";width
3.5423in;height 2.6403in;depth 0pt;original-width 8.8566in;original-height
6.602in;cropleft "0";croptop "1";cropright "1";cropbottom "0";filename
'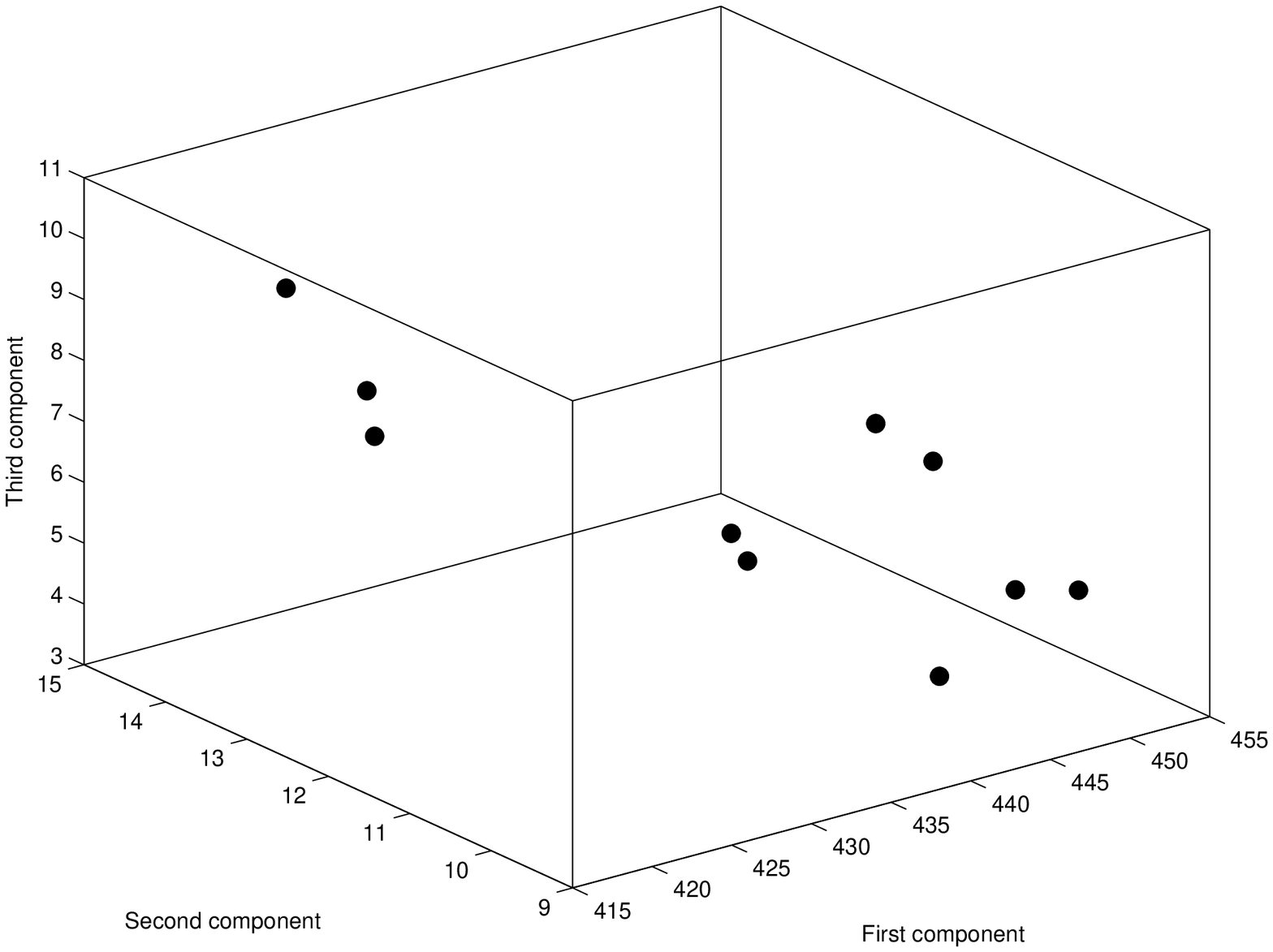';file-properties "XNPEU";}}

An analysis of individual countries also reveals interesting facts. The
scatter plot of the component scores (Fig.~\ref{fig:scores}) shows three
points that stand apart from the rest. The most extreme case, having the
smallest first-component score and the largest third-component score, is
Finland. This is an unexpected result for someone unfamiliar with Finnish
history, but it turns out that Finland was fighting on two different fronts
during the war years. A quick comparison of the individual mortality plots
(shown in the Technical Report) reveals that Finland, indeed, experienced
the largest increase in mortality rate for the 20-40 age bracket during the
war years among the countries in this sample (this is precisely what a small
first-component score accompanied by a large third-component score
indicates, according to our interpretation of the components).

The other two atypical points are Spain and Italy. Spain did not participate
in the Second World War but went through a civil war in the 1930s, showing a
different mortality pattern from the rest of the countries; in particular,
the decrease in child mortality after 1945 was not as fast as for the other
countries. Italy, by contrast, has the largest second-component score and is
the country with the fastest post-war decrease in infant mortality.

This example illustrates the kind of insight that can be gained from the
functional singular value decomposition. While other methods (like
tensor-product splines) can provide estimators of the mean function, the
FSVD also offers an interpretable decomposition of the mean that can reveal
interesting aspects of the data.

\section{Simulations\label{sec:Simulations}}

As mentioned before, we see the FSVD mainly as a tool for graphical and
exploratory data analysis, but since (\ref{eq:mu_p}) can be used as an
estimator of $\mu $, we ran a Monte Carlo study to compare its performance
with that of tensor-product spline estimators. Specifically, we wanted to
assess the ability of our free-knot component estimators to adapt to local
features of $\mu $, and the potential dangers of underestimating the
approximation order $p$.

We generated data from a mean-plus-error model $x_{ijk}=\mu
(s_{j},t_{k})+\varepsilon _{ijk}$. Two different means were considered, $\mu
_{1}(s,t)=\sum_{k=1}^{2}\lambda _{k}^{1/2}\phi _{k}(s)\psi _{k}(t)$ and $\mu
_{2}(s,t)=\sum_{k=1}^{3}\lambda _{k}^{1/2}\phi _{k}(s)\psi _{k}(t)$, with $%
\phi _{k}(s)=\sqrt{2}\sin (2k\pi s)$, $\psi _{k}(t)=\sqrt{2}\cos (2k\pi t)$, 
$\lambda _{1}=1$, $\lambda _{2}=1/2$ and $\lambda _{3}=1/32$. The grids $%
\{s_{j}\}$ and $\{t_{k}\}$ consisted of $m=r$ equispaced points in $[0,1]$,
and the errors $\varepsilon _{ijk}$ were independent $N(0,\sigma ^{2})$. We
considered two grid sizes, $m=20$ and $m=30$, two sample sizes, $n=10$ and $%
n=50$, and two error variances, $\sigma ^{2}=1$ and $\sigma ^{2}=4$. Each
model was replicated 200 times (although not all combinations of factors
were considered; see Table \ref{tab:simulation_errors}).

For the tensor-product spline estimator, we took two bases of cubic
B-splines with knots placed at the grid points. The estimator was
regularized by penalizing the integrated squared partial derivatives, as
explained in Hastie et al. (2001, ch. 5). The choice of a good smoothing
parameter is crucial for the behavior of these estimators. To be as fair as
possible with tensor-product splines, we chose the optimal smoothing
parameter: the minimizer of $\Vert \hat{\mu}-\mu \Vert $. In practice this
cannot be done because $\mu $ is unknown, so the estimation errors reported
in Table \ref{tab:simulation_errors} (under \textquotedblleft
TPS\textquotedblright ) will be lower than those attainable in practice.

As FSVD estimator of $\mu $ we took a two-component decomposition, $\hat{\mu}%
^{(2)}$, with $\hat{\phi}_{k}$s and $\hat{\psi}_{k}$s estimated by free-knot
cubic splines, as explained in Section \ref{sec:Estimation}. Here the number
of knots plays the role of smoothing parameter, so we considered two
possibilities: a fixed number of knots (3 for $\phi _{1}$, 5 for $\phi _{2}$%
, 2 for $\psi _{1}$ and 4 for $\psi _{2}$), and an optimal number of knots
(the number that minimizes $\Vert \hat{\phi}_{k}-\phi _{k}\Vert $ or $\Vert 
\hat{\psi}_{k}-\psi _{k}\Vert $, up to a maximum of 10 knots). The
estimation errors are reported in Table \ref{tab:simulation_errors} as
\textquotedblleft SVf\textquotedblright\ and \textquotedblleft
SVo\textquotedblright , respectively. These two are extreme cases, so the
actual estimation error of $\hat{\mu}^{(2)}$ when the number of knots is
selected by the user will fall somewhere between these two.

%TCIMACRO{\TeXButton{B}{\begin{table}[tbp] \centering}}%
%BeginExpansion
\begin{table}[tbp] \centering%
%EndExpansion
\begin{tabular}{lllllllllllll}
\hline
\multicolumn{7}{c}{Model parameters} &  & \multicolumn{5}{c}{Root ISE} \\ 
\cline{1-7}\cline{9-13}
Mean &  & $\sigma $ &  & $m$ &  & $n$ &  & TPS &  & SVf &  & SVo \\ \hline
&  &  &  &  &  &  &  &  &  &  &  &  \\ 
$\mu _{1}$ &  & 1 &  & 20 &  & 10 &  & .159 &  & .111 &  & .097 \\ 
&  &  &  &  &  & 50 &  & .085 &  & .075 &  & .047 \\ 
&  &  &  & 30 &  & 10 &  & .114 &  & .090 &  & .069 \\ 
&  &  &  &  &  & 50 &  & .063 &  & .070 &  & .034 \\ 
&  &  &  &  &  &  &  &  &  &  &  &  \\ 
$\mu _{1}$ &  & 2 &  & 20 &  & 10 &  & .277 &  & .196 &  & .184 \\ 
&  &  &  &  &  & 50 &  & .147 &  & .103 &  & .089 \\ 
&  &  &  & 30 &  & 10 &  & .197 &  & .140 &  & .124 \\ 
&  &  &  &  &  & 50 &  & .104 &  & .086 &  & .062 \\ 
&  &  &  &  &  &  &  &  &  &  &  &  \\ 
$\mu _{2}$ &  & 2 &  & 20 &  & 10 &  & .285 &  & .264 &  & .255 \\ 
&  &  &  &  &  & 50 &  & .160 &  & .205 &  & .197 \\ 
&  &  &  & 30 &  & 10 &  & .212 &  & .225 &  & .217 \\ 
&  &  &  &  &  & 50 &  & .110 &  & .196 &  & .187 \\ \hline
\end{tabular}%
\caption{Simulation Results. Root mean integrated squared errors for
tensor-product spline estimator (TPS) and FSVD estimators with fixed number
of knots (SVf) and optimal number of knots (SVo).}\label%
{tab:simulation_errors}%
%TCIMACRO{\TeXButton{E}{\end{table}}}%
%BeginExpansion
\end{table}%
%EndExpansion

Table \ref{tab:simulation_errors} shows the root integrated squared errors, $%
E^{1/2}(\left\Vert \hat{\mu}-\mu \right\Vert ^{2})$. Standard errors are not
given, to avoid overcrowding the table, but all the differences are
significant (the Technical Report shows boxplots of the simulated squared
errors). We see that for $\mu _{1}$, for which the order $p$ of $\hat{\mu}$
is correctly specified, the FSVD estimator with a fixed number of knots
outperforms the tensor-product spline estimator in all situations but one ($%
\sigma =1$, $m=30$, $n=50$), while the FSVD estimator with optimal number of
knots outperforms the tensor-product spline estimator in \emph{all}
situations (usually by a considerable margin).

For $\mu _{2}$ the situation reverses, as expected, since the order $p$ is
now underspecified and then the bias does not vanish, even as $m$ or $n$
increase. Of course, it can be argued that $p$ in practice is also chosen in
a data-driven way: for large $m$ and $n$, the estimators $\hat{\phi}_{3}$
and $\hat{\psi}_{3}$ will be regular enough to call for a three-component
estimator, which will make the FSVD estimator competitive again. The
conclusion of this Monte Carlo study, then, is that FSVD estimators are
competitive and even better than tensor-product splines as long as the
number of components is not severely underspecified. Even if the estimated
eigenvalues are small, for estimation purposes it is safer to include as
many eigenfunctions as possible, as long as they are not overwhelmed by
noise.

\section*{Acknowledgment}

This research was supported by the National Science Foundation, award number
DMS 0604396.

\appendix{}

\section{Appendix}

The following proofs use functional analysis results that can be found, for
instance, in Gohberg et al.~(2003). Given $\mu \in L^{2}(\mathcal{S}\times 
\mathcal{T})$, define the operator $\mathfrak{M}:L^{2}(\mathcal{T}%
)\rightarrow L^{2}(\mathcal{S})$ as $(\mathfrak{M}f)(s)=\int_{\mathcal{T}%
}\mu (s,t)f(t)\mathrm{d}t$. The adjoint of $\mathfrak{M}$ is the operator $%
\mathfrak{M}^{\ast }:L^{2}(\mathcal{S})\rightarrow L^{2}(\mathcal{T})$ given
by $(\mathfrak{M}^{\ast }g)(t)=\int_{\mathcal{S}}\mu (s,t)g(s)\mathrm{d}s$.
Let $\mathfrak{K}_{1}=\mathfrak{MM}^{\ast }$ and $\mathfrak{K}_{2}=\mathfrak{%
M}^{\ast }\mathfrak{M}$. They are self-adjoint operators, $\mathfrak{K}%
_{1}:L^{2}(\mathcal{S})\rightarrow L^{2}(\mathcal{S})$ and $\mathfrak{K}%
_{2}:L^{2}(\mathcal{T})\rightarrow L^{2}(\mathcal{T})$, with kernels $%
k_{1}(s_{1},s_{2})=\int \mu (s_{1},t)\mu (s_{2},t)\mathrm{d}t$ and $%
k_{2}(t_{1},t_{2})=\int \mu (s,t_{1})\mu (s,t_{2})\mathrm{d}s$, respectively.

Remember that for $f\in \mathcal{H}_{1}$ and $g\in \mathcal{H}_{2}$, the
tensor-product operator $g\otimes f:\mathcal{H}_{1}\rightarrow \mathcal{H}%
_{2}$ is defined as $(g\otimes f)(h)=\langle f,h\rangle g$.

\subsection{Proof of Theorem 1}

Since $\mathfrak{K}_{2}$ is a self-adjoint integral operator, the spectral
decomposition implies that $\mathfrak{K}_{2}=\sum \lambda _{k}\psi
_{k}\otimes \psi _{k}$, where $\lambda _{k}>0$ and $\{\psi _{k}\}$ is an
orthonormal system of eigenfunctions of $\mathfrak{K}_{2}$, which can be
completed to a basis of $L^{2}(\mathcal{T})$ by adding an orthonormal basis
of $\ker (\mathfrak{K}_{2})$, say $\{\tilde{\psi}_{k}\}$ (Gohberg et al.,
2003, p.~180). This proves (2) of Theorem 1. Note that $\ker (\mathfrak{K}%
_{2})=\ker (\mathfrak{M})$: clearly $\ker (\mathfrak{M})\subseteq \ker (%
\mathfrak{K}_{2})$ because $\mathfrak{K}_{2}=\mathfrak{M}^{\ast }\mathfrak{M}
$; but for any $f\in \ker (\mathfrak{K}_{2})$, $0=\langle f,\mathfrak{K}%
_{2}f\rangle =\Vert \mathfrak{M}f\Vert ^{2}$, which implies $f\in \ker (%
\mathfrak{M})$ and then $\ker (\mathfrak{K}_{2})\subseteq \ker (\mathfrak{M}%
) $.

Now define $\phi _{k}=\lambda _{k}^{-1/2}\mathfrak{M}\psi _{k}.$The $\phi
_{k}$s are orthonormal in $L^{2}(\mathcal{S})$, since 
\begin{eqnarray*}
\langle \phi _{j},\phi _{k}\rangle &=&\lambda _{j}^{-1/2}\lambda
_{k}^{-1/2}\langle \mathfrak{M}\psi _{j},\mathfrak{M}\psi _{k}\rangle \\
&=&\lambda _{j}^{-1/2}\lambda _{k}^{-1/2}\langle \psi _{j},\mathfrak{K}%
_{2}\psi _{k}\rangle =\lambda _{j}^{-1/2}\lambda _{k}^{-1/2}~\lambda
_{k}\delta _{jk}.
\end{eqnarray*}

To prove (3) of Theorem 1, define the operator $\mathfrak{L}=\sum \lambda
_{k}^{1/2}\phi _{k}\otimes \psi _{k}$. This operator is well defined, since
for any $f\in L^{2}(\mathcal{T})$, we have $\mathfrak{L}f=\sum \lambda
_{k}^{1/2}\langle \psi _{k},f\rangle \phi _{k}$ and 
\begin{equation*}
\Vert \mathfrak{L}f\Vert ^{2}=\sum \lambda _{k}|\langle \psi _{k},f\rangle
|^{2}\leq \Vert f\Vert ^{2}\sum \lambda _{k}<\infty .
\end{equation*}%
Direct calculation shows that $\mathfrak{L}\psi _{k}=\mathfrak{M}\psi _{k}$,
and $\mathfrak{L}\tilde{\psi}_{k}=\mathfrak{M}\tilde{\psi}_{k}=0$ because $%
\ker (\mathfrak{K}_{2})=\ker (\mathfrak{M})$. Since $\{\psi _{k}\}\cup \{%
\tilde{\psi}_{k}\}$ is a basis of $L^{2}(\mathcal{T})$, it follows that $%
\mathfrak{L}=\mathfrak{M}$, which is (3) of Theorem 1 in different words.

The identity (1) of Theorem 1 follows from (3), since $\mathfrak{K}_{1}=%
\mathfrak{MM}^{\ast }$. In particular, this shows that the positive
eigenvalues of $\mathfrak{K}_{1}$ are the same as those of $\mathfrak{K}_{2}$%
, and the $\phi _{k}$s can be taken as the corresponding eigenfunctions.

If the mean function $\mu (s,t)$ is continuous, Mercer's Theorem (Gohberg et
al., 2003, p.~198) implies that the $\psi _{k}$s are continuous and $k_{2}$
satisfies (2) in Theorem 1 in a pointwise manner, with the series converging
absolutely and uniformly.

The $\phi _{k}$s are continuous by definition when $\mu $ is continuous. To
prove that the identity (1) in Theorem 1 holds pointwise and that the series
converges absolutely and uniformly, we essentially mimic the proof of
Mercer's Theorem. See the Technical Report for details.

Finally, to show that expression (3) in Theorem 1 holds pointwise when the
series on the right-hand side converges absolutely and uniformly, note that
both sides of expression (3) define the same operator from $L^{2}(\mathcal{T}%
)$ to $L^{2}(\mathcal{S})$, so the identity must hold almost everywhere, and
by continuity, it must actually hold everywhere.$\blacksquare $

\bigskip

\textbf{Remark.} As by-products of the proof of Theorem 1 we obtain the
identities 
\begin{equation*}
\phi _{k}(s)=\frac{1}{\lambda _{k}^{1/2}}(\mathfrak{M}\psi _{k})(s)=\frac{1}{%
\lambda _{k}^{1/2}}\int \mu (s,t)\psi _{k}(t)\mathrm{d}t,
\end{equation*}%
and%
\begin{equation*}
\psi _{k}(t)=\frac{1}{\lambda _{k}^{1/2}}(\mathfrak{M}^{\ast }\phi _{k})(t)=%
\frac{1}{\lambda _{k}^{1/2}}\int \mu (s,t)\phi _{k}(s)\mathrm{d}s.
\end{equation*}

\subsection{Proof of Theorem 2}

Since $\{f_{k}\}$ and $\{g_{k}\}$ are orthonormal, 
\begin{equation*}
\left\Vert \mu -h\right\Vert ^{2}=\left\Vert \mu \right\Vert
^{2}-2\sum_{k=1}^{p}a_{k}\langle g_{k},\mathfrak{M}f_{k}\rangle
+\sum_{k=1}^{p}a_{k}^{2},
\end{equation*}%
which is minimized by $a_{k}=\langle g_{k},\mathfrak{M}f_{k}\rangle ,\
k=1,\ldots ,p$. Then, minimizing $\left\Vert \mu -h\right\Vert ^{2}$ is
equivalent to maximizing $\sum_{k=1}^{p}\left\vert \langle g_{k},\mathfrak{M}%
f_{k}\rangle \right\vert ^{2}$. By Cauchy-Schwartz inequality, 
\begin{eqnarray}
\sum_{k=1}^{p}\left\vert \langle g_{k},\mathfrak{M}f_{k}\rangle \right\vert
^{2} &\leq &\sum_{k=1}^{p}\left\Vert g_{k}\right\Vert ^{2}\left\Vert 
\mathfrak{M}f_{k}\right\Vert ^{2}  \notag \\
&=&\sum_{k=1}^{p}\left\vert \langle \mathfrak{M}f_{k},\mathfrak{M}%
f_{k}\rangle \right\vert ^{2}=\sum_{k=1}^{p}\left\vert \langle f_{k},%
\mathfrak{K}_{2}f_{k}\rangle \right\vert ^{2}.  \label{eq:upper_bound}
\end{eqnarray}%
It is well known (or see Gohberg et al., 2003, Section 4.9) that (\ref%
{eq:upper_bound}) is maximized by the leading $p$ eigenfunctions of $%
\mathfrak{K}_{2}$, and the maximum value is $\sum_{k=1}^{p}\lambda _{k}$.
Therefore $\sum_{k=1}^{p}\left\vert \langle g_{k},\mathfrak{M}f_{k}\rangle
\right\vert ^{2}\leq \sum_{k=1}^{p}\lambda _{k}$ and equality holds for $%
f_{k}=\psi _{k}$ and $g_{k}=\phi _{k}$, which completes the proof. $%
\blacksquare $

\subsection{Proof of Theorem 3}

Let $z_{ijk}=x_{ijk}-\mu (s_{j},t_{k})$, and define $\mathbf{M}_{0}=[\mu
(s_{j},t_{k})]_{(j,k)}$, $\mathbf{X}_{i}=[x_{ijk}]_{(j,k)}$ and $\mathbf{Z}%
_{i}=[z_{ijk}]_{(j,k)}$. Since $\mathbf{\hat{\Omega}=B}^{\top }\mathbf{VK}%
_{1}\mathbf{VB}$ and $\mathbf{K}_{1}\mathbf{=\bar{X}U\bar{X}}^{\top }$, we
can write 
\begin{eqnarray}
\hat{\Omega}_{hh^{\prime }} &=&\beta _{h}(\mathbf{s})^{\top }\mathbf{V\bar{X}%
U\bar{X}}^{\top }\mathbf{V}\beta _{h^{\prime }}(\mathbf{s})  \notag \\
&=&\beta _{h}(\mathbf{s})^{\top }\mathbf{VM}_{0}\mathbf{UM}_{0}^{\top }%
\mathbf{V}\beta _{h^{\prime }}(\mathbf{s})  \label{eq:Omega_1} \\
&&+2\beta _{h}(\mathbf{s})^{\top }\mathbf{V\bar{Z}UM}_{0}^{\top }\mathbf{V}%
\beta _{h^{\prime }}(\mathbf{s})  \label{eq:Omega_2} \\
&&+\beta _{h}(\mathbf{s})^{\top }\mathbf{V\bar{Z}U\bar{Z}}^{\top }\mathbf{V}%
\beta _{h^{\prime }}(\mathbf{s}).  \label{eq:Omega_3}
\end{eqnarray}%
We will show that (\ref{eq:Omega_1}) goes to $\Omega _{hh^{\prime }}$ as $m$
and $r$ go to infinity, and that (\ref{eq:Omega_2}) and (\ref{eq:Omega_3})
go to zero in probability as $n$ goes to infinity, uniformly in $m$ and $r$.

Since 
\begin{equation*}
\beta _{h}(\mathbf{s})^{\top }\mathbf{V\bar{X}U\bar{X}}^{\top }\mathbf{V}%
\beta _{h^{\prime }}(\mathbf{s})=
\end{equation*}%
\begin{equation*}
\sum_{j=1}^{m}\sum_{j^{\prime }=1}^{m}\beta _{h}(s_{j})v_{j}\left\{
\sum_{k=1}^{r}u_{k}\mu (s_{j},t_{k})\mu (s_{j^{\prime }},t_{k})\right\}
v_{j^{\prime }}\beta _{h^{\prime }}(s_{j^{\prime }}),
\end{equation*}
it is clear that (\ref{eq:Omega_1}) goes to $\Omega _{hh^{\prime }}$ as $m$
and $r$ go to infinity, because both $\max v_{j}$ and $\max u_{k}$ go to
zero as $m$ and $r$ go to infinity.

With respect to (\ref{eq:Omega_2}), note that we can write it as $2\bar{y}$,
with 
\begin{equation*}
y_{i}=\beta _{h}(\mathbf{s})^{\top }\mathbf{VZ}_{i}\mathbf{UM}_{0}^{\top }%
\mathbf{V}\beta _{h^{\prime }}(\mathbf{s}).
\end{equation*}
The $y_{i}$s are i.i.d.~with $\mathrm{E}(y_{i})=0$ and $\mathrm{V}(y_{i})=%
\mathrm{V}\left\{ \sum_{j=1}^{m}\sum_{k=1}^{r}\beta
_{h}(s_{j})v_{j}z_{ijk}u_{k}a_{kh^{\prime }}\right\} $, with $a_{kh^{\prime
}}=\sum_{j^{\prime }=1}^{m}\mu (s_{j^{\prime }},t_{k})v_{j^{\prime }}\beta
_{h^{\prime }}(s_{j^{\prime }})$. It can be proved that 
\begin{equation*}
\lim_{\substack{ m\rightarrow \infty  \\ r\rightarrow \infty }}\mathrm{V}%
(y_{i})=\iiiint \beta _{h}(s_{1})\alpha _{h^{\prime }}(t_{1})\beta
_{h}(s_{2})\alpha _{h^{\prime }}(t_{2})\rho \{(s_{1},t_{1}),(s_{2},t_{2})\}%
\mathrm{d}s_{1}\mathrm{d}s_{2}\mathrm{d}t_{1}\mathrm{d}t_{2},
\end{equation*}%
where $\alpha _{h^{\prime }}(t_{k})=\int \mu (s,t_{k})\beta _{h^{\prime }}(s)%
\mathrm{d}s$ as $m\rightarrow \infty $ (see Technical Report). Then $\mathrm{%
V}(y_{i})$ is bounded for any $m$ and $r$, and a simple application of
Tchebyshev's Inequality implies that (\ref{eq:Omega_2}) goes to zero in
probability as $n$ goes to infinity, uniformly in $m$ and $r$.

Regarding (\ref{eq:Omega_3}), note that 
\begin{equation*}
\beta _{h}(\mathbf{s})^{\top }\mathbf{V\bar{Z}U\bar{Z}}^{\top }\mathbf{V}%
\beta _{h^{\prime }}(\mathbf{s})\leq \Vert \mathbf{U}^{1/2}\mathbf{\bar{Z}}%
^{\top }\mathbf{V}\beta _{h}(\mathbf{s})\Vert \Vert \mathbf{U}^{1/2}\mathbf{%
\bar{Z}}^{\top }\mathbf{V}\beta _{h^{\prime }}(\mathbf{s})\Vert .
\end{equation*}%
For a given index $h$, we can write $\mathbf{U}^{1/2}\mathbf{\bar{Z}}^{\top }%
\mathbf{V}\beta _{h}(\mathbf{s})=\mathbf{\bar{w}}$, with $\mathbf{w}_{i}=%
\mathbf{U}^{1/2}\mathbf{Z}_{i}^{\top }\mathbf{V}\beta _{h}(\mathbf{s})$. The 
$\mathbf{w}_{i}$s are i.i.d.~with $\mathrm{E}(\mathbf{w}_{i})=0$ and 
\begin{equation*}
\lim_{\substack{ m\rightarrow \infty  \\ r\rightarrow \infty }}\sum_{k=1}^{r}%
\mathrm{V}(w_{ik})=\iiint \beta _{h}(s_{1})\beta _{h}(s_{2})\rho
\{(s_{1},t),(s_{2},t)\}\mathrm{d}s_{1}\mathrm{d}s_{2}\mathrm{d}t
\end{equation*}%
(again, see Technical Report). Since $\mathrm{E}(\Vert \mathbf{\bar{w}}\Vert
^{2})=n^{-1}\sum_{k=1}^{r}\mathrm{V}(w_{ik})$, a straightforward application
of Markov's Inequality implies that $\Vert \mathbf{\bar{w}}\Vert $ goes to
zero in probability as $n$ goes to infinity, uniformly in $m$ and $r$, and
consequently the same is true for (\ref{eq:Omega_3}). $\blacksquare $

\section*{References}

\begin{description}
\item Adler, R. J., and Taylor, J. E. (2007). \emph{Random Fields and
Geometry}. New York: Springer-Verlag.

\item Aubry, N., Guyonnet, R., and Lima, R. (1991). Spatio-temporal analysis
of complex signals: theory and applications.\ \emph{Journal of Statistical
Physics}, 64, 683--739.

\item De Boor, C. (2001). \emph{A Practical Guide to Splines}. New York:
Springer-Verlag.

\item Dente, J. A., Vilela Mendes, R., Lambert, A., and Lima, R. (1996). The
bi-orthogonal decomposition in image processing: signal analysis and texture
segmentation.\ \emph{Signal Processing: Image Communication}, 8, 131--148.

\item Ferraty, F., and Vieu, P. (2006). \emph{Nonparametric Functional Data
Analysis: Theory and Practice}. New York: Springer-Verlag.

\item Gasser, T., Gervini, D., and Molinari, L. (2004). Kernel estimation,
shape-invariant modeling and structural analysis.\ In \emph{Methods in Human
Growth Research}, eds. R. Hauspie, N. Cameron \& L. Molinari. Cambridge:
Cambridge University Press, pp.179--204.

\item Gervini, D. (2006). Free-knot spline smoothing for functional data.\ 
\emph{Journal of the Royal Statistical Society}, Ser. B, 68, 671--687.

\item Gohberg, I., Goldberg, S., and Kaashoek, M. A. (2003). \emph{Basic
Classes of Linear Operators}. Basel: Birkh\"{a}user Verlag.

\item Gu, C. (2000). Multivariate spline regression. In \emph{Smoothing and
Regression: Approaches, Computation, and Application}, ed. M. Schimek. New
York: Wiley, pp. 329--355.

\item H\"{a}rdle, W., and M\"{u}ller, M. (2000). Multivariate and
semiparametric kernel regression.\ In \emph{Smoothing and Regression:
Approaches, Computation, and Application}, ed. M. Schimek. New York: Wiley,
pp. 357--391.

\item Hastie, T., Tibshirani, R., and Friedman, J. (2001). \emph{The
Elements of Statistical Learning. Data Mining, Inference and Prediction.}
New York: Springer.

\item Kneip, A., and Utikal, K. J. (2001). Inference for density families
using functional principal component analysis.\ \emph{Journal of the
American Statistical Association}, 96, 519--532.

\item Morris, J. S., and Carroll, R. J. (2006). Wavelet-based functional
mixed models.\ \emph{Journal of the Royal Statistical Society}, Ser. B, 68,
179--199.

\item Nychka, D. (2000). Spatial-process estimates as smoothers.\ In \emph{%
Smoothing and Regression: Approaches, Computation, and Application}, ed. M.
Schimek. New York: Wiley, pp. 393--424.

\item Ramsay, J. O., and Silverman, B. W. (2002). \emph{Applied Functional
Data Analysis: Methods and Case Studies}. New York: Springer-Verlag.

\item Ramsay, J. O., and Silverman, B. W. (2005). \emph{Functional Data
Analysis} (2nd edition). New York: Springer-Verlag.

\item Silverman, B. W. (1996). Smoothed functional principal components
analysis by choice of norm.\ \emph{Annals of Statistics}, 24, 1--24.

\item Taylor, J. E., and Worsley, K. J. (2007). Detecting sparse signals in
random fields, with an application to brain mapping.\ \emph{Journal of the
American Statistical Association}, 102, 913--928.

\item Tyler, D. E. (1981). Asymptotic inference for eigenvectors.\ \emph{The
Annals of Statistics}, 9, 725--736.

\item Yao, F., and Lee, T. C. M. (2006). Penalized spline models for
functional principal component analysis.\ \emph{Journal of the Royal
Statistical Society}, Ser. B, 68, 3--25.
\end{description}

\end{document}